\documentclass[3p,times]{elsarticle}

\usepackage{ecrc}

\volume{00}

\firstpage{1}

\journalname{Nuclear Physics A}

\runauth{}

\jid{npa}

\jnltitlelogo{Nuclear Physics A}

\usepackage{amssymb}

\biboptions{square,comma,numbers,sort&compress}

\usepackage[figuresright]{rotating}

\newcommand{\Slash}[1]{{\ooalign{\hfil/\hfil\crcr$#1$}}}

\begin{document}
\begin{frontmatter}



\dochead{}

\title{Di-jet asymmetric momentum transported by QGP fluid}

\author{Y.~Tachibana\corref{cor1}\fnref{label1,label2,label3}}
\ead{tachibana@nt.phys.s.u-tokyo.ac.jp}
\author[label3]{T.~Hirano}
\ead{hirano@sophia.ac.jp}

\address[label1]{Department of Physics, The University of Tokyo, Tokyo 113-0033, Japan}
\address[label2]{Theoretical Research Division, Nishina Center, RIKEN, Wako 351-0198, Japan}
\address[label3]{Department of Physics, Sophia University, Tokyo 102-8554, Japan}

\address{}

\begin{abstract}
We study the collective flow of the {QGP}-fluid 
which transports the energy and momentum 
deposited from jets. 
Simulations of the propagation of jets 
together with expansion of the {QGP}-fluid 
are performed 
by solving relativistic hydrodynamic equations 
numerically 
in the fully $\left(3+1\right)$-dimensional space. 
Mach cones are induced by 
the energy-momentum deposition from jets 
and extended by the expansion of the {QGP}. 
As a result, 
low-$p_T$ particles are enhanced 
at large angles from the jet axis. 
This provedes an intimate link between 
the observables in di-jet asymmetric events 
in heavy ion collisions 
and 
theoretical pictures of 
the medium excitation by jet-energy deposition. 

\end{abstract}


\end{frontmatter}



\section{Introduction}
In heavy-ion collisions at 
Relativistic Heavy Ion Collider ({RHIC}) in {BNL} 
and Large Hadron Collider ({LHC}) in {CERN}, 
the deconfined phase of quarks and gluons, 
namely the quark-gluon plasma ({QGP})
is supposed to be realized experimentally. 
The expansion of the {QGP} 
is well described by 
relativistic hydrodynamics 
\cite{Heinz:2001xi, sQGP1, sQGP2, sQGP3, Hirano:2005wx}.
At the same time as the {QGP}, 
large-$p_{T}$ partons,
so-called jets, 
are produced through 
the initial hard scattering between the partons inside 
colliding nuclei and penetrate the {QGP}. 
Due to the strong coupling with the {QGP}-medium, 
these jet particles travel losing their energy. 
At the leading order, 
the jet particles are created as 
a back-to-back pair 
with the same energy 
owing to the energy-momentum conservation law. 
Depending on the relation between 
the geometry of the medium and 
the position of the pair creation,
the amount of the energy loss differs 
between the jet partons. 
At the {LHC}, events 
such that the transverse momentum of jets 
are highly-asymmetric
are observed 
and these experimental facts are 
consistent with jet quenching picture 
\cite{Aad:2010bu, Chatrchyan:2011sx}. 
Furthermore, 
according to the data from the {CMS} Collaboration, 
this imbalance of dijet-$p_{T}$ is 
compensated by 
low-$p_{T}$ particles 
at large angles from the jet axis 
and 
the total-$p_{T}$ of the entire system 
is well balanced \cite{Chatrchyan:2011sx}. 
It can be supposed that 
these low-$p_{T}$ particles 
are originating from 
a medium wake induced by 
the energy and momentum 
deposited from jets. 

Here, we perform simulations of 
dijet asymmetric events 
to study the transport process of 
energy and momentum deposited 
from jet particles in the expanding QGP medium. 
To describe the medium response 
to the energy-momentum deposition, 
we solve relativistic hydrodynamic equations 
with source terms 
in the fully (3+1)-dimensional coordinate system 
numerically. 
Then, we investigate 
the transverse-momentum balance in dijet events 
and 
show that
low-$p_{T}$ particles 
at large angles from the jet axis 
play a crucial role 
in the transverse-momentum balance. 
\section{Model and Simulations}
Assuming local thermal equilibrium of the {QGP}, 
we solve relativistic hydrodynamic equations, 
to describe the space-time evolution of the {QGP}. 
Here, source terms, which are the 4-momentum density 
deposited from the traversing jet partons, 
are introduced in the hydrodynamic equations.
\begin{eqnarray}
\partial_{\mu}T^{\mu \nu}\left(x\right)=J^{\nu}\left(x\right), \label{eqn:hydro_source}
\end{eqnarray}
where $T^{\mu \nu}$ and $J^{\nu}$ are 
the energy-momentum tensor of the {QGP}-fluid and
the source terms, 
respectively. 
The energy-momentum tensor for perfect fluids 
can be decomposed 
using the 4-flow velocity $u^{\mu}=\gamma \left( 1,{\mbox{\boldmath $v$}}\right)$ as 
\begin{eqnarray}
T^{\mu \nu}&=&(\epsilon+P)u^{\mu}u^{\nu}-Pg^{\mu \nu}. 
\end{eqnarray}
Here $\epsilon$ is the energy density, 
$P$ is the pressure and 
$g^{\mu\nu}={\rm diag}\left(1, -1, -1, -1\right)$ 
is the Minkowski metric. 
As an equation of state, the ideal gas equation of state 
for massless partons 
is employed here: $P\left(\epsilon\right)=\epsilon /3$. 
We assume that 
the deposited energy and momentum 
are immediately equilibrated. 
The source terms for 
a massless particle traveling through 
the fluid are given by 
\begin{eqnarray}
J^{\mu}
&=&
-
\frac{dp_{\rm jet}^0}{dt}  
\frac{p_{\rm jet}^{\mu}}{p_{\rm jet}^0} 
\delta^{(3)}\left(\mbox{\boldmath $x$}-\mbox{\boldmath $x$}_{\rm jet}(t)\right). 
\end{eqnarray}
For $dp_{\rm jet}^0/dt$, the collisional energy loss is used 
\cite{Thoma:1990fm}. 
Here we multiply the energy loss by a constant, 
whose value is fixed throughout all simulations, 
to simulate large asymmetric dijet events 
such as observed at the LHC. 
We solve the equations (\ref{eqn:hydro_source}) numerically 
without linearization 
in the $\left(3+1\right)$-dimensional Milne coordinates 
$\left(\tau, x, y, \eta_{s}\right)$. 
This framework enables us to simulate 
the collective flow induced by jet particles 
on 
an expanding medium. 
Through the Cooper-Frye formula, 
we calculate the momentum distribution of particles 
from hydrodynamic outputs 
$T\left(x\right),\:u^{\mu}\left(x\right)$ 
\cite{Cooper:1974mv}. 

We set up the initial profile of the {QGP}-fluid 
at $\tau_0 = 0.6\:{\rm fm}/c$. 
The initial energy density in the $\eta_{s}$-direction 
is flat like the Bjorken scaling solution 
in the mid-rapidity region $\left|\eta_{s}\right|<10$. 
The flat region is smoothly connected to the vacuum 
at both ends by a half Gaussian with the width 
$\sigma_{\eta}$=0.5 
\cite{Bjorken:1982qr, Schenke:2011tv}. 
For the initial transverse profile, 
we use the smooth energy density 
for central Pb-Pb collisions 
obtained from the MC-Glauber model 
\cite{Hirano:2010je}. 
The initial value of energy density 
at origin 
is 
$e\left(\tau_0,\:\mbox{\boldmath $x$}=0\right)=100\:{\rm GeV/fm^3}$. 
A back-to-back pair of massless partons 
is supposed to be produced 
at $\left(\tau=0, x=x_0, y=0, \eta_{s}=0\right)$ 
with the same energy, 
$p^{0}_{\rm jet} (\tau=0) = 200\:{\rm GeV}$. 
These partons always travel 
along the $x$-axis 
at the speed of light 
and start to interact with the expanding medium 
at $\tau=\tau_0$. 
To characterize 
the $p_{T}$-imbalance 
between the pair of the partons, 
the asymmetry ratio is introduced. 
\begin{eqnarray}
A_{J} = \frac{p_{T,1}-p_{T,2}}{p_{T,1}+p_{T,2}}, 
\end{eqnarray}
where $p_{T,1}$ and $p_{T,2}$ are 
transverse momenta of the energetic partons 
at the freeze-out ($p_{T,1}>p_{T,2}$). 
The value of $A_{J}$ 
can be controlled by 
moving the position of pair creation 
$x_0$. 

\section{Results}
\begin{figure}[tbp]
\vspace*{8mm}
\begin{tabular}{cc}
\begin{minipage}{0.5\hsize}
\begin{center}
\includegraphics[]{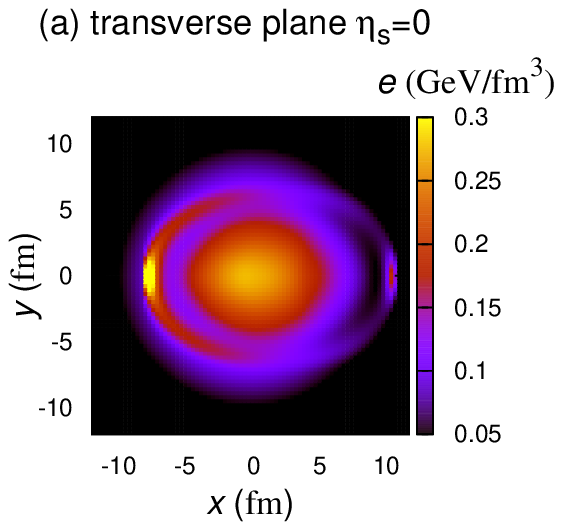}
\end{center}
\end{minipage}
\begin{minipage}{0.5\hsize}
\begin{center}
\includegraphics[]{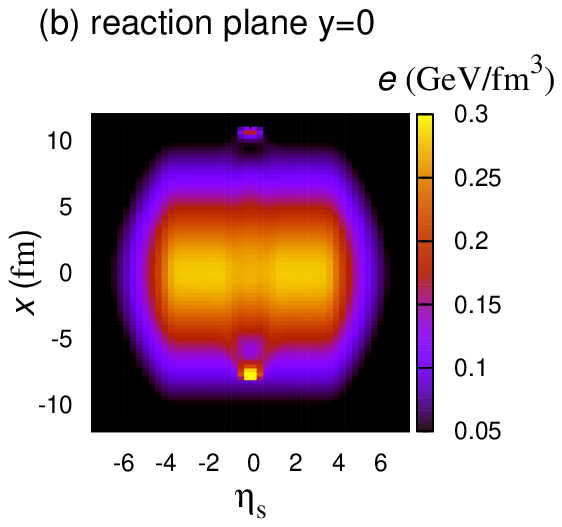}
\end{center}
\end{minipage}
\end{tabular}
 \caption{(Colour Online)
Energy density distribution of the {QGP}-fluid at 
$\tau=0.6\:{\rm fm}/c$ in the transverse plane at 
$\eta_{s}=0$ (a) and 
in the reaction plane at $y=0$ (b). 
The pair-production position of the energetic partons is 
$\left(\tau=0, x=1.5\:{\rm fm}, y=0, \eta_{s}=0\right)$. 
The energetic partons travel in the opposite direction 
along the $x$-axis at the speed of light. 
}
\label{fig:map}
\end{figure}

The energy density distribution 
of the {QGP} fluid at $\tau=9.6\:{\rm fm}/c$ 
is shown in Figure \ref{fig:map}. 
The pair of the energetic partons is 
produced at $x_0=1.5\:{\rm fm}$ in this case. 
One sees 
oval structures 
of higher energy density, 
which are Mach cones 
\cite{Stoecker:2004qu, CasalderreySolana:2004qm}
distorted by the radial flow of the fluid, 
in the transverse plane at $\eta_{s}=0$ 
(Fig. \ref{fig:map} (a)). 
In the reaction plane $y=0$ (Fig.~\ref{fig:map} (b)) 
Mach cone-like structures are also seen. 
Though they do not develop so much apparently 
because the Milne coordinate itself expands 
in the longitudinal direction. 

\begin{figure}[h]
\begin{center}
\vspace*{7mm}
\includegraphics[width=0.8 \linewidth]{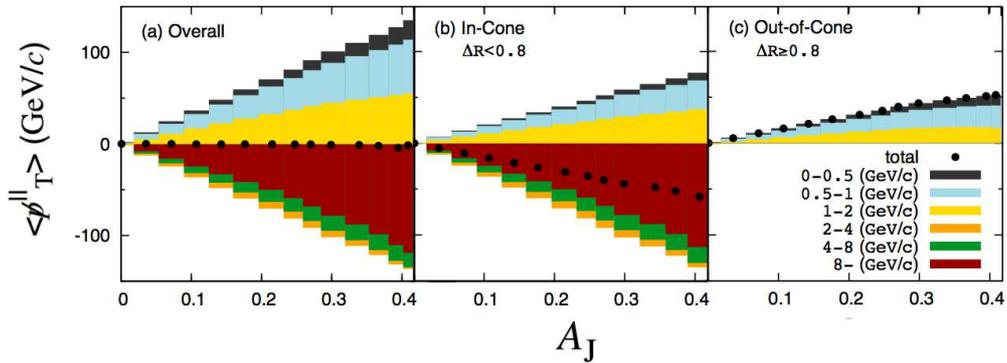}
\end{center}
 \caption{(Colour Online)
Total-$p_{T}$ along the jet axis. 
$\langle \protect\Slash{p}^{||}_{T} \rangle$ 
is shown as a function of $A_{J}$ 
for the whole region (a), 
for inside  ($\Delta R < 0.8$) jet cones (b) 
and for outside ($\Delta R \geq 0.8$) jet cones (c). 
The solid black circles indicate 
the contribution to $\langle \protect\Slash{p}^{||}_{T} \rangle$ 
from the particles in the whole $p_{T}$ region. 
Each coloured histogram corresponds to the contribution 
from the particles 
in six transverse momentum ranges: 
$0$-$0.5$, 
$0.5$-$1$, 
$1$-$2$, 
$2$-$4$, 
$4$-$8$ GeV/$c$, 
and $p_{T}>8\:{\rm GeV}/c$. 
}
\label{fig:ptpar}
\end{figure}

To see the 
total $p_{T}$-balance 
of the entire system, 
total-$p_{T}$ along the jet axis 
is defined as 
\begin{eqnarray}
\langle \Slash{p}^{||}_{T} \rangle = \langle \sum_{i}-p^{i}_{T}\cos(\phi_{i}-\phi_{1}) \rangle,
\end{eqnarray}
where the sum is taken over 
all particles in the entire system, 
and their $p_{T}$ is projected onto 
the sub-leading jet axis $\phi_2 = \phi_1 + \pi$
in the transverse plane. 
We calculate 
the contribution to 
$\langle \Slash{p}^{||}_{T} \rangle$ 
from the fluid through 
the Cooper-Frye formula \cite{Cooper:1974mv}. 
We assume that the freeze-out occurs at fixed proper time 
$\tau_{f}=9.6\:{\rm fm}/c$, 
which is typical for central Pb-Pb collisions at the {LHC} energy. 
By adding the contribution from  the jet particles
to the contribution from the fluid, 
we obtain $\langle \Slash{p}^{||}_{T} \rangle$. 
Figure \ref{fig:ptpar} 
shows $\langle \Slash{p}^{||}_{T} \rangle$ 
as a function of $A_{J}$. 
Here, we consider 2-jet cones 
of the size 
$\Delta R = \sqrt{\Delta \phi^2 + \Delta \eta^2}=0.8$ 
around the axes of the energetic partons. 
Figures \ref{fig:ptpar} (a), (b) and (c) 
show the results for 
overall, in-cone and out-of-cone,
respectively. 
Each coloured histogram corresponds to 
the contribution from 
a $p_{T}$-region of the particles. 
The solid black circles show
the contribution from 
the whole particles. 
In the in-cone region, 
the contribution from high-$p_{T}$ particles 
are dominant and negative. 
On the other hand, 
there is only 
the positive contribution from 
particles of $p_{T}<2\:{\rm GeV}/c$ 
in the out-of-cone region. 
These contributions 
from the in-cone region 
and from the out-of-cone region 
are balanced 
due to the energy-momentum conservation 
of the entire system. 
The low-$p_{T}$ particles 
in the out-of-cone region 
are originated from 
the deposited energy and momentum 
transported 
by the collective flow in the {QGP}-fluid. 
\section{Summary}
We reported that 
low-$p_{T}$ particles 
are enhanced 
at large angles 
from the jet axes 
as a result of 
the transport 
of the energy and momentum 
deposited from jets 
in the expanding {QGP} medium. 
We performed simulations of
asymmetric dijet events 
in Pb-Pb collisions at the LHC. 
To describe the collective flow 
induced by the energy-momentum deposition, 
we solved relativistic hydrodynamic equations 
with source terms 
numerically 
in the fully $\left(3+1\right)$-dimensional space. 
Mach cones are excited 
by the energy deposited from jets 
and then distorted 
by the expansion of the background medium. 
We also found that 
low-$p_{T}$ particles are enhanced at large angles 
from the jet axes 
and compensate a large fraction of 
the transverse momentum 
deposited from jets. 
This low-$p_{T}$ enhancement at large angles 
is caused by 
transport 
of the deposited energy and momentum 
by the collective flow in the {QGP}-fluid. 
This study 
suggests 
an intimate link between 
the data from {CMS} 
and 
medium response to 
energy-momentum deposition of jets.

\section*{Acknowledgments}
The work of Y.~T. is supported by 
a JSPS Research Fellowship for Young Scientists 
(KAKENHI Grant Numbers 13J02554) 
and by an Advanced Leading Graduate Course for Photon Science. 
The work of T.~H. is supported by 
JSPS KAKENHI Grant Numbers 25400269. 








\end{document}